\documentclass[12pt]{iopart}

\usepackage{iopams}

\usepackage{graphicx}% Include figure files

\def\U#1{{%
\def\O{\mbox{O}}
\def\u{\mbox{u}}
\mathcode`\u=\mu
\mathcode`\O=\Omega
\mathrm{#1}}}

\def\degree{\mbox{$^\circ$}}

\def\ii{{\mathrm{i}}}

\def\ee{{\mathrm{e}}}

\def\dd{{\mathrm{d}}}

\def\Re{\mathop{\mathrm{Re}}}

\def\Im{\mathop{\mathrm{Im}}}

\def\bra#1{\langle #1|}

\def\ket#1{|#1\rangle}

\def\bracket#1{\langle#1\rangle}

\def\sub#1{_{\mathrm{#1}}}

\def\sur#1{^{\mathrm{#1}}}

\newcommand{\nn}{\nonumber \\}
\newcommand{\abs}[1]{\left\vert#1\right\vert}

\DeclareMathSymbol{\varGamma}{\mathord}{letters}{"00}%"
\DeclareMathSymbol{\varDelta}{\mathord}{letters}{"01}%"
\DeclareMathSymbol{\varTheta}{\mathord}{letters}{"02}%"
\DeclareMathSymbol{\varLambda}{\mathord}{letters}{"03}%"
\DeclareMathSymbol{\varXi}{\mathord}{letters}{"04}%"
\DeclareMathSymbol{\varPi}{\mathord}{letters}{"05}%"
\DeclareMathSymbol{\varSigma}{\mathord}{letters}{"06}%"
\DeclareMathSymbol{\varUpsilon}{\mathord}{letters}{"07}%"
\DeclareMathSymbol{\varPhi}{\mathord}{letters}{"08}%"
\DeclareMathSymbol{\varPsi}{\mathord}{letters}{"09}%"
\DeclareMathSymbol{\varOmega}{\mathord}{letters}{"0A}%"

\begin{document}

\title{Weak measurements with completely mixed probe states}

\author{S Tamate, T Nakanishi and M Kitano}%

\address{Department of Electronic Science and Engineering, Kyoto University, Kyoto 615-8510, Japan}

\ead{tamate@giga.kuee.kyoto-u.ac.jp}

\begin{abstract}
Weak measurements with mixed probe states were investigated in the presence of
noise in the probe system.
We show that the completely mixed state can be used as a probe state
for weak measurements of imaginary weak values.
The completely mixed probe state has some advantages in terms of noise tolerance.
We also experimentally demonstrated weak measurements of polarization rotation
via unpolarized light, which corresponds to weak measurements with completely mixed probe states
in a two-state system.
\end{abstract}

% Uncomment for PACS numbers title message

\maketitle

\section{\label{sec:level1}Introduction}

A basic tenet of quantum mechanics is that the state of quantum system describes only
its statistical properties.
A state can be pre-selected by a projective measurement and
its properties can be investigated by repeating measurements on identically
prepared ensembles of the system.
In their seminal work, Aharonov, Bergmann, and Lebowitz \cite{Aharonov1964} proposed
further splitting the pre-selected ensemble into subensembles based on the result of a second projective measurement; this process is known as post-selection.
The pre- and post-selected ensemble contains statistical information regarding the state between the two measurements.
This way of describing a system by pre- and post-selected states
is known as the two-state vector formalism \cite{Aharonov2007a}.

Aharonov, Albert, and Vaidman \cite{Aharonov1988} proposed using weak measurements
(an indirect measurements with weakening of the interaction)
to measure the expectation value of an observable in a pre- and post-selected ensemble.
They found that the ensemble average of an observable $\hat{A}$ is characterized
by the so-called weak value
\begin{equation}
 \bracket{\hat{A}}\sub{w} = \frac{\bracket{\psi\sub{f}|\hat{A}|\psi\sub{i}}}{\bracket{\psi\sub{f}|\psi\sub{i}}},  \label{eq:1}
\end{equation}
where $\ket{\psi\sub{i}}$ and $\ket{\psi\sub{f}}$ represent pre- and post-selected states, respectively.
Unlike usual expectation values for pre-selected-only ensembles,
weak values can be anomalously large and exceed the range of eigenvalues.
This means that a particular pre- and post-selected ensemble may have the potential to considerably affect other systems (or probes) through even a weak interaction.
Large weak values have been experimentally observed in various optical systems \cite{Ritchie1991,Parks1998,Solli2004,Brunner2004,Pryde2005,Wang2006,Iinuma2011}.

Potentially large weak values hold out the possibility of ``amplifying'' the effect of weak interactions by appropriately designing the pre- and post-selected ensemble.
The ``amplification'' scheme using pre- and post-selection is known as weak-value amplification.
In 2008, Hosten and Kwiat demonstrated the usefulness of weak-value amplification by observing the spin Hall effect of light via weak measurements \cite{Hosten2008}.
Inspired by this experiment, weak-value amplification has been applied
to various precision measurements including
beam deflection measurements \cite{Dixon2009, Hogan2011, Turner2011}, frequency measurements \cite{Starling2010}, and measurements of the plasmonic spin Hall effect \cite{Gorodetski2012}.

Measuring the effect of feeble interactions requires performing many runs of an experiment.
The post-selection in the weak-value amplification is designed to
extract only the subensemble that has a high contribution to the signal;
the rest of the pre-selected ensemble is discarded.
The signal of weak measurements is enhanced when post-selection succeeds,
but the success probability decreases for larger enhancement factors.
Consequently, weak measurements are ultimately limited by the standard quantum limit.
Nevertheless, weak measurements are known to improve the
signal-to-noise ratio (SNR) of the experiments that are subject to noise.
Whether the SNR can be improved depends on the noise characteristics.
We assume that the dominant noise is proportional to $N^\alpha$, where $N$ denotes the total number of experimental runs and $\alpha$ is determined by factors such as the noise statistics and measurement method.
For $\alpha > 1/2$, weak measurements produce a net gain in the SNR
by a factor of $\mathcal{P}(\mathrm{f|i})^{1/2-\alpha}$, where $\mathcal{P}(\mathrm{f|i})$
is the success probability of the post-selection \cite{Hogan2011}.
More detailed analyses have been given for specific noise such as noise in optical experiments \cite{Starling2009,Kobayashi2011a} and correlated noise \cite{Feizpour2011}, both of which correspond to the case $\alpha = 1$.
Since practical experiments can have
multiple noise sources that have different values of $\alpha$,
weak values should be designed to suit the individual experiments \cite{Hogan2011}.

The weak value is generally complex and its real and imaginary parts can be measured
using the respective settings of the probe system. There are thus two possibilities when
designing a weak-value experiment.
A theoretical study indicates that using imaginary part of weak values for weak-value amplification
is robust to some kinds of noise in the initial probe states \cite{Kedem2012}.
This implies that a mixed, or noisy, probe state can be used for weak-value amplification.
As such an experiment, interferometric phase estimation via weak measurements with white light has been proposed \cite{Li2011a}.

The first aim of this study is to provide a unified view of weak measurements with
mixed probe states, and hence clarify the kinds of noise weak measurements can tolerate.
We demonstrate the possibility of weak measurements with completely mixed probe states.
It turns out that the completely mixed probe state has several advantageous properties for precision measurements.
The second aim of this study is to experimentally demonstrate weak measurements with completely mixed probe states.

The rest of the paper is organized as follows.
In Sec.~\ref{sec:level2}, we describe weak measurements with mixed probe states
by focusing mainly on weak measurements with imaginary weak values.
We also consider the robustness of the weak measurements to noise in the probe system.
In Sec.~\ref{sec:level3}, we demonstrate measurement of polarization rotation via unpolarized light, which corresponds to a weak measurement with a completely mixed probe state.
Section~\ref{sec:level4} summarizes the findings of the study.

\section{\label{sec:level2}Weak measurements with mixed probe states}

In this section, we describe the principle of weak measurements with mixed probe states
and discuss the noise tolerance of weak measurements.

\subsection{Fundamentals}

\begin{figure}
 \centering
 \includegraphics[width=250pt]{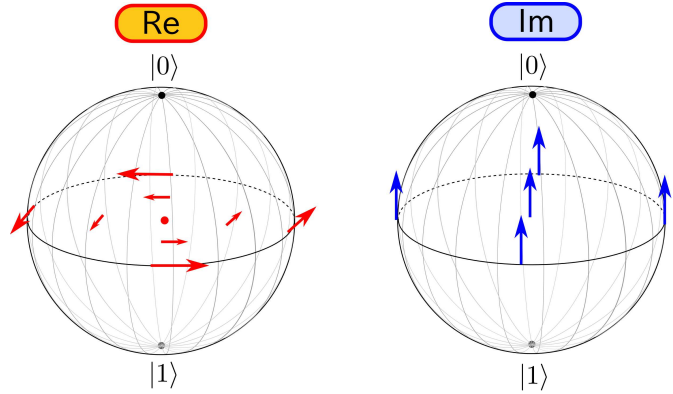}
 \caption{Flow of probe states induced by the real and imaginary parts of a weak value for $\hat{K} = \hat{Z}$ and the initial probe states with $\bracket{\hat{Z}}=0$ (on the equatorial plane of the Bloch ball).
The lengths of the arrows are proportional to the real and imaginary parts of the weak value.
The real and imaginary parts of the weak value respectively contribute to unitary and non-unitary evolutions of the probe states. The unitary flow becomes smaller as the probe state becomes more mixed. In contrast, the non-unitary flow is uniform in the equatorial plane.}
 \label{fig:1}
\end{figure}

Weak measurements involve two quantum systems: the measured and probe systems.
The measured system state is pre-selected in an initial state $\ket{\psi\sub{i}}$ and post-selected in a final state $\ket{\psi\sub{f}}$.
The ensemble is measured between pre- and post-selections via the following unitary evolution:
\begin{equation}
 \hat{U}(\theta) = \exp(-\ii \theta \hat{A}\otimes \hat{K}),  \label{eq:2}
\end{equation}
where $\hat{A}$ and $\hat{K}$ are respectively observables of the measured and probe systems and $\theta$ represents the strength of measurements.
The effective evolution of the probe system for the given pre- and post-selected ensemble is described by
\begin{equation}
 \hat{U}\sub{eff}(\theta) = \exp(-\ii \theta \bracket{\hat{A}}\sub{w} \hat{K}) + O(\theta^2),  \label{eq:3}
\end{equation}
which is derived from $\hat{U}(\theta)$ as
\begin{equation}
 \bra{\psi\sub{f}}\hat{U}(\theta)\ket{\psi\sub{i}} = \ee^{\ii \arg \bracket{\psi\sub{f}|\psi\sub{i}}}\sqrt{\mathcal{P}(\mathrm{f|i})}\hat{U}\sub{eff}(\theta),  \label{eq:4}
\end{equation}
where $\mathcal{P}(\mathrm{f|i}) \equiv \abs{\bracket{\psi\sub{f}|\psi\sub{i}}}^2 $ represents the success probability of post-selection.
Figure~\ref{fig:1} shows the effective evolution of a qubit probe with $\hat{K} = \hat{Z}$, where $\hat{Z} = \ket{0}\bra{0} - \ket{1}\bra{1}$ is the Pauli $Z$ operator.
The imaginary part of the weak value $\bracket{\hat{A}}\sub{w}$ contributes to the non-unitary evolution
and directly changes the probability distribution of $\hat{K}$,
while the real part contributes to the relative phase change of the probe states.

We prepare the probe state in a mixed state $\hat{\sigma}\sub{i}$. The final state of the probe is given by $\hat{\sigma}\sub{f}(\theta) = \mathcal{P}(\mathrm{f|i})\hat{U}\sub{eff}(\theta)\hat{\sigma}\sub{i}\hat{U}\sub{eff}(\theta)^\dagger$.
We denote the initial and final expectation values of a probe observable $\hat{M}$ by
$\bracket{\hat{M}}\sub{i} \equiv \tr(\hat{\sigma}\sub{i}\hat{M})/\tr\hat{\sigma}\sub{i}$ and $\bracket{\hat{M}}\sub{f} \equiv \tr[\hat{\sigma}\sub{f}(\theta)\hat{M}]/\tr\hat{\sigma}\sub{f}(\theta)$, respectively. We also define the shift operators for $\hat{M}$ as $\delta\sub{i}\hat{M} \equiv \hat{M} - \bracket{\hat{M}}\sub{i}$ and $\delta\sub{f}\hat{M} \equiv \hat{M} - \bracket{\hat{M}}\sub{f}$.
The shift of the probe observable $\hat{M}$ can be derived in a similar manner as the pure-state case \cite{Jozsa2007}:
\begin{equation}
 \bracket{\delta\sub{i}\hat{M}}\sub{f}
  = \theta \Re \bracket{A}\sub{w} \bracket{\ii[\hat{K}, \hat{M}]}\sub{i}
     + \theta \Im \bracket{A}\sub{w} \bracket{\{\delta\sub{i}\hat{K}, \delta\sub{i}\hat{M}\}}\sub{i}
     + O(\theta^2),  \label{eq:5}
\end{equation}
where $[\hat{A},\hat{B}] \equiv \hat{A}\hat{B} - \hat{B}\hat{A}$ and $\{\hat{A}, \hat{B}\} \equiv \hat{A}\hat{B} + \hat{B}\hat{A}$.
The factor $\bracket{\{\delta\sub{i}\hat{K}, \delta\sub{i}\hat{M}\}}\sub{i}$ represents the correlation
between $\hat{K}$ and $\hat{M}$ for the initial probe state.
The imaginary part of the weak value affects the probe observable correlated with $\hat{K}$.
Especially for $\hat{M} = \hat{K}$,
\begin{equation}
 \bracket{\delta\sub{i} \hat{K}}\sub{f}
  = 2 \theta \Im \bracket{A}\sub{w} \bracket{(\delta\sub{i} \hat{K})^2}\sub{i} + O(\theta^2).  \label{eq:6}
\end{equation}
We can extract the contribution of only the imaginary part of the weak value by measuring $\hat{K}$.
Hereafter, we treat weak measurements with $\hat{M} = \hat{K}$.

We consider the SNR of the measurement.
The final variance of $\hat{K}$ is given by $\bracket{(\delta\sub{f}\hat{K})^2}\sub{f} = \bracket{(\delta\sub{i}\hat{K})^2}\sub{i} + O(\theta)$ (for details, see \ref{sec:appendixA}) and the success probability of post-selection is $\tr\hat{\sigma}\sub{f} = \mathcal{P}(\mathrm{f|i}) + O(\theta)$.
Repeating the measurement $N$ times, the post-selection succeeds $N\tr\sigma\sub{f}$ times on average.
Hence, the SNR of the measurement is derived as
\begin{eqnarray}
 \mathrm{SNR} &= \frac{N\tr\hat{\sigma\sub{f}}\,\bracket{\delta\sub{i} \hat{K}}\sub{f}}{\sqrt{N\tr\hat{\sigma\sub{f}}\,\bracket{(\delta\sub{f} \hat{K})^2}\sub{f}}} \nn
  &=  2 \theta \Im \bracket{A}\sub{w} \sqrt{N\mathcal{P}(\mathrm{f|i})\bracket{(\delta\sub{i} \hat{K})^2}\sub{i}} + O(\theta^2). \label{eq:7}
\end{eqnarray}
The SNR is proportional to the standard deviation $\sqrt{\bracket{(\delta\sub{i} \hat{K})^2}\sub{i}}$ of the initial probe state; which is independent of the coherence between eigenstates $\ket{k}$ of $\hat{K}$.
Therefore, the decoherence of the initial probe state does not hinder measurement provided the standard deviation is retained.
The completely mixed state can also be used as an initial probe state.
Especially for a qubit probe, the completely mixed state always achieves the maximum SNR since it has the maximum standard deviation for any $\hat{K}$.
This is in contrast to the real part of the weak value, which cannot be measured by using the completely mixed state since $\ii[\hat{K}, \hat{M}]$ in Eq.~(\ref{eq:5}) is a traceless operator.

\subsection{Noise tolerance}

\begin{figure}
 \centering
 \includegraphics[width=350pt]{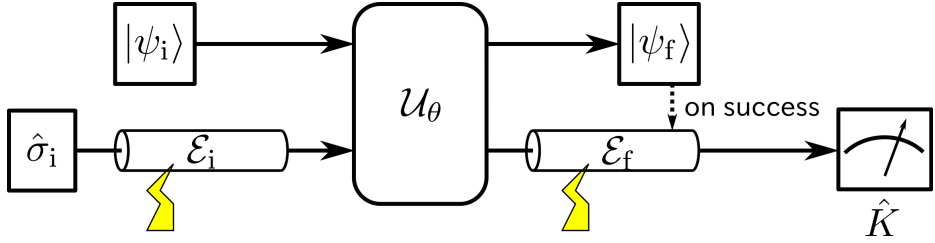}
 \caption{Setup for weak measurements with mixed probe states.
 The probe system is exposed to the noises $\mathcal{E}\sub{i}$ and $\mathcal{E}\sub{f}$
 before and after the measurement interaction.}
 \label{fig:2}
\end{figure}

We describe the noise tolerance of the weak measurements in detail.
Figure~\ref{fig:2} shows a schematic diagram of our setup.
We assume that the probe system is exposed to the noises before and after the interaction, which are expressed by the quantum channels $\mathcal{E}\sub{i}$ and $\mathcal{E}\sub{f}$, respectively.

For convenience, we refer to quantum channels $\mathcal{E}$ that satisfy the following condition as phase noise. For an arbitrary eigenstate $\ket{k}$ of $\hat{K}$,
\begin{equation}
  \mathcal{E}(\ket{k}\bra{k}) = \ket{k}\bra{k}.  \label{eq:8}
\end{equation}
This condition is satisfied if
and only if the quantum channel has the Kraus representation $\{\hat{E}_n\}$ of the form
\begin{equation}
 \hat{E}_n = \sum_k c_n(k)\ket{k}\bra{k},  \label{eq:9}
\end{equation}
where $c_n(k)$ are complex numbers satisfying $\sum_{n}\abs{c_n(k)}^2 = 1$.
When the probe system is a two-state system and $\hat{K} = \hat{Z}$,
it can be easily verified that the phase noise is simply the
composition of phase-flip noise and a unitary rotation about the $Z$ axis.

We show that if $\mathcal{E}\sub{i}$ and $\mathcal{E}\sub{f}$ are phase noises
the result of weak measurements is unaffected by these noises.
We define a map representing the measurement interaction as $\mathcal{U}_\theta(\hat{\rho}) \equiv \hat{U}(\theta)\hat{\rho}\hat{U}(\theta)^\dagger$, where $\hat{\rho}$ is a state of the whole system.
Assuming that $\mathcal{E}$ is the phase noise, Eq.~(\ref{eq:9}) indicates that
$\mathcal{E}$ has the following two properties:
\begin{eqnarray}
 (\mathcal{I}\otimes\mathcal{E}) \circ \mathcal{U}_\theta = \mathcal{U}_\theta \circ (\mathcal{I}\otimes\mathcal{E}),  \label{eq:10}\\
 \bra{k}\mathcal{E}(\hat{\sigma})\ket{k} = \bra{k}\hat{\sigma}\ket{k},  \label{eq:11}
\end{eqnarray}
where $\mathcal{I}$ is the identity channel for the measured system and $\hat{\sigma}$ is an arbitrary probe state.
It directly follows from Eq.~(\ref{eq:8}) that the composition of phase noises is also a phase noise.
Let $p'\sub{f}(k)$ and $p\sub{f}(k)$ respectively denote the final probability distributions of the weak measurement with and without noise. Then
\begin{eqnarray}
 p'\sub{f}(k) &= \bra{\psi\sub{f}}\bra{k}(\mathcal{I} \otimes \mathcal{E}\sub{f})\circ \mathcal{U}_\theta \circ(\mathcal{I} \otimes \mathcal{E}\sub{i})(\ket{\psi\sub{i}}\bra{\psi\sub{i}}\otimes\hat{\sigma\sub{i}})\ket{\psi\sub{f}}\ket{k} \nn
  &= \bra{\psi\sub{f}}\bra{k} (\mathcal{I} \otimes \mathcal{E}\sub{f} \circ \mathcal{E}\sub{i})\circ\mathcal{U}_\theta(\ket{\psi\sub{i}}\bra{\psi\sub{i}}\otimes\hat{\sigma\sub{i}})\ket{\psi\sub{f}}\ket{k} \nn
  &= \bra{\psi\sub{f}}\bra{k} \mathcal{U}_\theta(\ket{\psi\sub{i}}\bra{\psi\sub{i}}\otimes\hat{\sigma\sub{i}})\ket{\psi\sub{f}}\ket{k}
  = p\sub{f}(k).  \label{eq:12}
\end{eqnarray}
Hence, the results of weak measurements are unaffected by the phase noises $\mathcal{E}\sub{i}$ and $\mathcal{E}\sub{f}$.

In addition, using the completely mixed probe state has a further advantage in terms of noise tolerance.
In fact, any noise described by a unital channel, which maps the identity operator $\hat{I}$ to itself (i.e. $\mathcal{E}\sub{i}(\hat{I}) = \hat{I}$) cannot affect the completely mixed state.
Therefore, the result of weak measurements will be insensitive to a wider class of noise before the measurement interaction.

The weak measurement of the imaginary part of the weak value seems to be more
classical than that of the real part
because the probe observable $\hat{K}$ commutes with the measurement interaction.
Interestingly, this classicality contributes to robustness against probe noise.

\section{Experiments\label{sec:level3}}

\begin{figure}[t]
 \centering
 \includegraphics[width=400pt]{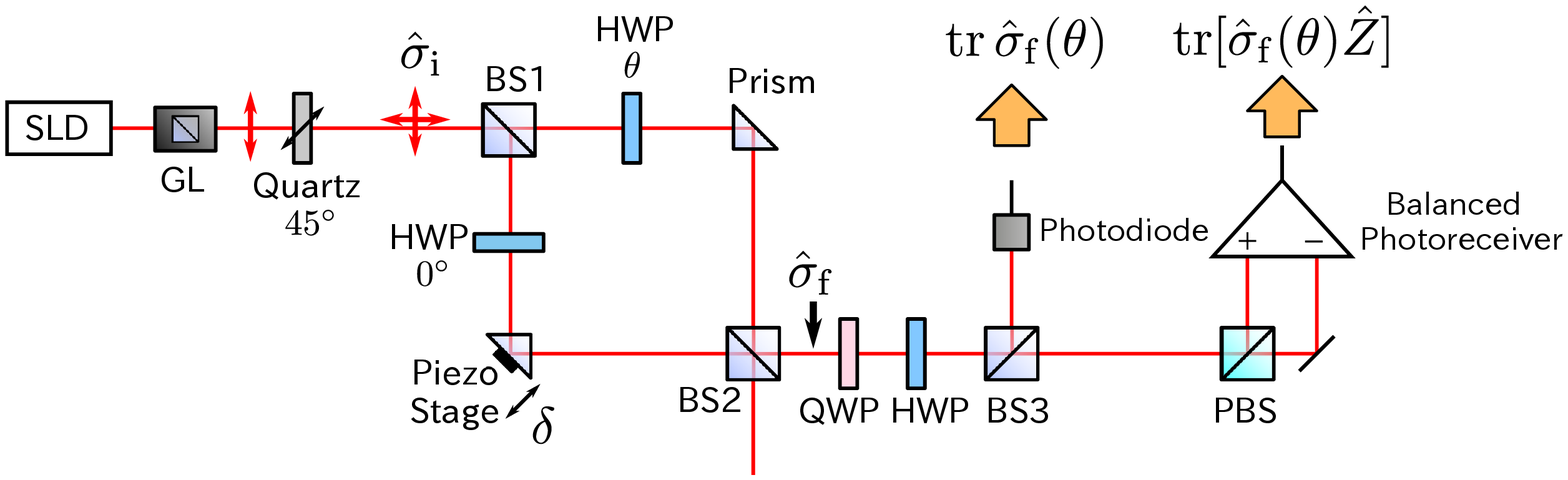}
 \caption{Experimental setup for weak measurements of polarization rotation via unpolarized light.
The unpolarized light is produced by using a superluminescent diode (SLD), a Glan laser polarizer (GL), and a $5$-$\U{mm}$-wide quartz plate.
The SLD (QSDM-780-9, Qphotonics) has a center wavelength of $785\,\U{nm}$ and a spectral width of $17.3\,\U{nm}$. The spatial mode of the SLD is filtered by a single-mode fiber.
The polarization of the output light is projected onto the vertically polarized state by the GL, and then depolarized by the quartz plate.
Since the differential group delay in the quartz plate is larger than the coherence time of the light, unpolarized light can be obtained by aligning the optical axis of the quartz plate at $45\degree$.
The optical power of the unpolarized light is about $1\,\U{mW}$.
The unpolarized light is introduced to a Mach--Zehnder interferometer. The prism in the lower path is mounted on a piezo stage. Each arm of the interferometer has a half-wave plate (HWP). The interaction strength between the path and the polarization can be varied by rotating the optical axis $\theta$ of the HWP in the upper path. The power and polarization of the light are measured at one of the output ports. The quarter-wave plate (QWP) and HWP are adjusted to measure the circularly polarized components of the light.}
 \label{fig:3}
\end{figure}

In this section, we experimentally demonstrate weak measurements with completely mixed probe states.
Figure~\ref{fig:3} shows the setup used for these experiments.
The measured and probe systems correspond to the paths and polarization of photons, respectively.
We used the unpolarized light as a completely mixed probe state and thereby
measured the path-dependent polarization rotation via weak measurements.

To generate the unpolarized light, we used a superluminescent diode (SLD), a Glan laser polarizer (GL), and a quartz plate.
The SLD output has a sufficiently short coherence time to be depolarized by
differential group delay in the quartz plate.
The first beam splitter (BS1) pre-selects the path state as
\begin{equation}
 \ket{\psi\sub{i}} = \frac{1}{\sqrt{2}}(\ket{0} + \ket{1}),  \label{eq:13}
\end{equation}
where $\ket{0}$ and $\ket{1}$ respectively represents the upper and lower path states of the Mach--Zehnder interferometer. Post-selection of the path was realized by observing one of the output ports of BS2. We can change
the relative phase $\delta$ of the post-selected state $\ket{\psi\sub{f}}$ by translating the piezo stage in the lower path:
\begin{equation}
 \ket{\psi\sub{f}} = \frac{1}{\sqrt{2}}(\ket{0} + \ee^{\ii\delta} \ket{1}).  \label{eq:14}
\end{equation}
The interaction between the measured and probe systems was implemented by inserting a half-wave plate (HWP) in each arm. The optical axis of the upper HWP was rotated by an angle $\theta$. In practical applications, $\theta$ corresponds to an unknown physical parameter to be estimated.
Let $\hat{U}\sub{HWP(\theta)}$ denote the unitary operation caused by the HWP at angle $\theta$. The unitary evolution of the whole system is then given by
\begin{eqnarray}
 \hat{U}(\theta) &= \ket{0}\bra{0}\otimes \hat{U}\sub{HWP(\theta)} + \ket{1}\bra{1}\otimes\hat{U}\sub{HWP(0^\circ)} \nn
  &= \hat{U}\sub{HWP(0^\circ)} \left[\ket{0}\bra{0}\otimes\exp(2 \ii \theta \hat{Z}) +\ket{1}\bra{1}\otimes \hat{I} \right],  \label{eq:15}
\end{eqnarray}
where $\hat{Z}$ is an observable distinguishing right- and left-handed circular polarization.
We can eliminate the overall polarization rotation $\hat{U}\sub{HWP(0^\circ)}$ by adjusting the measurement basis for polarization.
Therefore, we regard the unitary evolution in the Mach--Zehnder interferometer as
\begin{equation}
 \hat{U}(\theta) = \exp \left( 2 \ii \theta \hat{P}_0 \otimes \hat{Z} \right),  \label{eq:16}
\end{equation}
where we denote the projector onto the upper path state as $\hat{P}_0 \equiv \ket{0}\bra{0}$.
The weak value of the upper path operator $\hat{P}_0$ is calculated as
\begin{equation}
 \Im \bracket{\hat{P}_0}\sub{w} = \frac{1}{2} \tan \left(\frac{\delta}{2}\right),  \label{eq:17}
\end{equation}
which diverges at $\delta = \pi$.

\begin{figure}
 \centering
 \includegraphics[width=300pt]{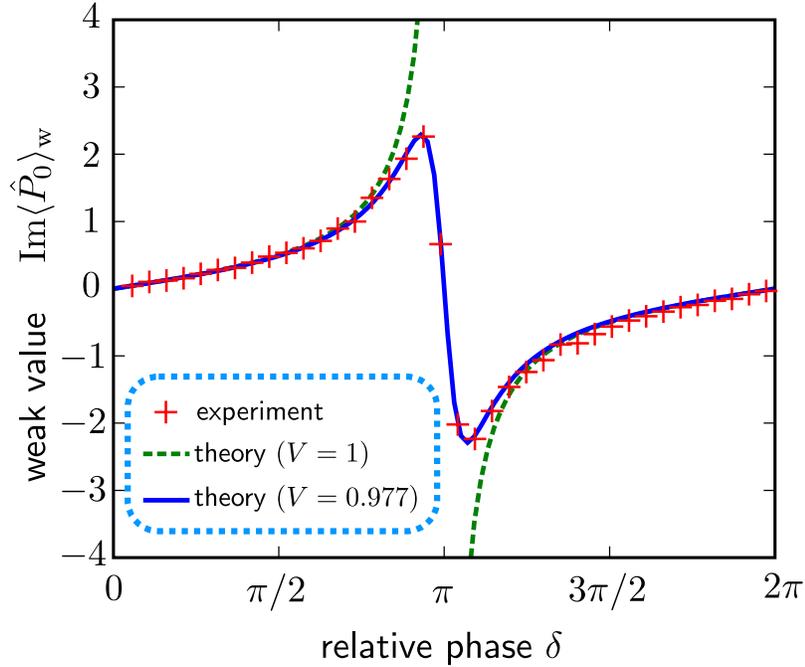}
 \caption{Weak values for various post-selected states. The crosses represent the experimental data. The dashed green curve and the solid blue curve represent the theoretical curves for $V = 1$ and $V = 0.977$, respectively.}
 \label{fig:4}
\end{figure}

At the output of the interferometer, we measured the success probability of the post-selection and
the circular components of the polarization.
Recalling that the initial probe state is unpolarized ($\hat{\sigma}\sub{i} = \hat{I}/2$), the expected result is given by
\begin{eqnarray}
 \tr\hat{\sigma}\sub{f}(\theta) = \frac{1}{2}(1 + \cos\delta \cos(2\theta)),  \label{eq:18}\\
 \tr[\hat{\sigma}\sub{f}(\theta)\hat{Z}] = -\frac{1}{2}\sin\delta \sin(2\theta).  \label{eq:19}
\end{eqnarray}
The imaginary part of the weak value can be extracted from these values since
it corresponds to the normalized sensitivity to the parameter $\theta$:
\begin{equation}
 \Im \bracket{\hat{P}_0} = -\frac{1}{4} \left.\frac{\dd}{\dd \theta}\right|_{\theta = 0} \frac{\tr[\hat{\sigma}\sub{f}(\theta)\hat{Z}]}{\tr\hat{\sigma}\sub{f}(\theta)}.  \label{eq:20}
\end{equation}
We obtained the gradient at $\theta = 0$ by linearly fitting the normalized results of the polarization measurements in the range $-2\degree \le \theta \le 2\degree$.
We derived the weak value from the gradient.
Repeating this procedure for various $\delta$, we obtained
the weak values shown in Fig.~{\ref{fig:4}}.
The relative phases $\delta$ were estimated from the interference fringes; however, it had
an undesired offset due to the drift of the piezo stage during the experimental runs.
For this reason, the experimental data in Fig.~{\ref{fig:4}} is
translated horizontally to fit the theoretical curve.

The maximum weak value was $\Im\bracket{\hat{P}_0}\sub{w} = 2.26$. This is not very large
because post-selection of the path state was incomplete.
The actual post-selected state $\hat{\rho}\sub{f}$ is expressed in terms of the visibility $V$ for the case when there is no interaction ($\theta = 0$):
\begin{equation}
 \hat{\rho}\sub{f} = V\ket{\psi\sub{f}}\bra{\psi\sub{f}} + (1-V)\frac{\hat{I}}{2}.  \label{eq:21}
\end{equation}
The experimentally measured visibility was $V = 0.977$.
The weak value for mixed pre- and post-selected states has been derived by Wu and M\o{}lmer \cite{Wu2009} as
\begin{equation}
 \bracket{\hat{A}}\sub{w} = \frac{\tr(\hat{\rho}\sub{f}\hat{A}\hat{\rho}\sub{i})}{\tr(\hat{\rho}\sub{i}\hat{\rho}\sub{f})},  \label{eq:22}
\end{equation}
where $\hat{\rho}\sub{i}$ and $\hat{\rho}\sub{f}$ denote the pre- and post-selected states, respectively.
From Eq.~(\ref{eq:22}), we can calculate the weak value for incomplete post-selection as
\begin{equation}
 \Im \bracket{\hat{P}_0}\sub{w} = \frac{V\sin\delta}{2(1+V\cos\delta)}.  \label{eq:23}
\end{equation}
When $V = 1$, this equation is equivalent to Eq.~(\ref{eq:17}). The theoretical curve for the measured visibility $V=0.977$ agrees well with the experimental data, as shown in Fig.~\ref{fig:4}.

\section{Summary\label{sec:level4}}

We considered weak measurements with mixed probe states and demonstrated the advantages of weak measurements with imaginary weak values,
which tolerate phase noise in the probe system.
The completely mixed state has some advantageous properties for precision measurements.
We also experimentally demonstrated weak measurements with completely mixed probe states by measuring the polarization rotation via unpolarized light.
The unpolarized light itself is insensitive to polarization rotation; however, by attaching the path degree of freedom and using a weak measurement, unpolarized light can be used to measure polarization rotation.
Weak measurements with imaginary weak values are useful for designing highly sensitive measurements with poor probe states.
Weak measurements have the potential to open up new doors for measurements in practical noisy systems.

\ack
We thank Hirokazu Kobayashi for interesting and inspiring discussion.
One of the authors (S.T.) is supported by a JSPS Research Fellowship for Young Scientists (No. 224850).

\appendix

\section{General formula for the cumulant} \label{sec:appendixA}

We describe a general formula for calculating
the shift of the cumulant of $\hat{K}$ in weak measurements.
We introduce the cumulant generating function
\begin{equation}
 \varPhi(s) \equiv \log\bracket{\ee^{s\hat{K}}}
  = \log \frac{\tr(\hat{\sigma}\ee^{s\hat{K}})}{\tr\hat{\sigma}},  \label{eq:24}
\end{equation}
for the probe state $\hat{\sigma}$.
The $n$th-order cumulant $\bracket{\hat{K}^n}\sur{c}$ is
calculated from $\varPhi(s)$ as
\begin{equation}
 \bracket{\hat{K}^n}\sur{c} = \left.\frac{\dd^n \varPhi(s)}{\dd s^n}\right|_{s=0}.  \label{eq:25}
\end{equation}
A straightforward calculation gives the relationship between the cumulant generating functions $\varPhi\sub{i}(s)$ and $\varPhi\sub{f}(s)$ for the initial state $\hat{\sigma}\sub{i}$ and the final state $\hat{\sigma}\sub{f}$:
\begin{equation}
 \varPhi\sub{f}(s) = \varPhi\sub{i}(s + 2\theta\Im\bracket{\hat{A}}\sub{w}) - 2\theta\Im\bracket{\hat{A}}\sub{w}\bracket{\hat{K}}\sub{i} + O(\theta^2). \label{eq:26}
\end{equation}
This relation gives the general formula for calculating
the cumulant for the final state:
\begin{eqnarray}
 \bracket{\hat{K}^n}\sub{f}\sur{c} &= \left.\frac{\dd^n \varPhi\sub{f}(s)}{\dd s^n}\right|_{s=0} \nn
  &= \left.\frac{\dd^n \varPhi\sub{i}(s+2\theta\Im\bracket{\hat{A}}\sub{w})}{\dd s^n}\right|_{s=0} + O(\theta^2) \nn
  &= \left.\frac{\dd^n \varPhi\sub{i}(s)}{\dd s^n}\right|_{s=0} + 2\theta\Im\bracket{\hat{A}}\sub{w} \left.\frac{\dd^{n+1} \varPhi\sub{i}(s)}{\dd s^{n+1}}\right|_{s=0} + O(\theta^2) \nn
  &= \bracket{\hat{K}^n}\sub{i}\sur{c} + 2\theta\Im\bracket{A}\sub{w}\bracket{\hat{K}^{n+1}}\sub{i}\sur{c} + O(\theta^2).  \label{eq:27}
\end{eqnarray}
The case of $n=1$ corresponds to Eq.~(\ref{eq:6}), and the case of $n=2$ gives the
change in variance:
\begin{equation}
 \bracket{\delta\sub{f}\hat{K}^2}\sub{f} = \bracket{\delta\sub{i}\hat{K}^2}\sub{i} + 2\theta\Im\bracket{\hat{A}}\sub{w}\bracket{\hat{K}^3}\sub{i}\sur{c} + O(\theta^2).  \label{eq:28}
\end{equation}

\end{document}